\documentclass{elsart5p}
\usepackage{graphicx}
\usepackage{amsmath}
\usepackage{amssymb}
\begin{document}
\begin{frontmatter}
\title{Entanglement in the F-AF zig-zag Heisenberg chain}
\author{A. Avella},
\author{F. Mancini}, and
\author{E. Plekhanov}
\address{Dipartimento di Fisica ``E.R. Caianiello'' - Unit\`{a}
CNISM di Salerno \protect\\ Universit\`{a}
degli Studi di Salerno, I-84081 Baronissi (SA), Italy}
\begin{abstract}
We present a study of the entanglement properties of the F-AF zig-zag Heisenberg chain done by means of the Density Matrix Renormalization Group method. In particular, we have selected the concurrence as measure of entanglement and checked its capability to signal the presence of quantum phase transitions within the previously found ergodicity phase diagram [E. Plekhanov, A. Avella, and F. Mancini, Phys. Rev. B \textbf{74}, 115120 (2006)]. By analyzing the behavior of the concurrence, we have been able not only to determine the position of the transition lines within the phase diagram of the system, but also to identify a well defined region in the parameter space of the model that shows a complex spin ordering indicating the presence of a new phase of the system.
\end{abstract}
\end{frontmatter}

\section{Introduction}

Can we use any measure of entanglement as a tool to signal the presence of a quantum phase transition, and identify its nature, within the phase diagram of a physical system? Unfortunately, there is still no definitive answer to this question although more and more results, recently published in the literature, give clear evidences that this should be the case \cite{review}. But why do we care about using the properties of entanglement of a system to this end? The answer to this question exists, and is much simpler and immediate: almost any measure of entanglement can be almost exactly (after a finite-size scaling analysis) and automatically computed by means of numerical techniques (Lanczos, quantum Monte Carlo, Density Matrix Renormalization Group (DMRG)) and depends on a large number of correlations functions in a non-trivial manner. According to this, people hope to sketch the phase diagram of complex physical systems just analyzing the behavior of a single, easily computable, physical quantity instead of struggling to figure out which is the correlation function reporting the signature of a specific transition. On the other hand, the versatility of entanglement measures has a clear drawback: it is rather difficult to deeply comprehend the nature of a transition just looking at the behavior of such quantities in its proximity. At the end of the day, one has often to resort to correlation and response functions in order to classify, both in terms of nature and order, a transition. Therefore, we can just hope to use entanglement measures as cheap tools to position all transition lines over the phase diagram. At any rate, this should not be considered as a little achievement. As a matter of fact, after such a preliminary analysis, we could focus on few lines over a phase diagram instead of being forced to study the whole parameter space.

In this manuscript, we have studied the entanglement properties of the one-dimensional Heisenberg model with both nearest-neighbor and next-nearest-neighbor interactions. In particular, we have chosen a ferromagnetic z-axis nearest-neighbor interaction and an antiferromagnetic in-plane nearest-neighbor interaction. The next-nearest-neighbor interaction is antiferromagnetic and isotropic. The anisotropy in the nearest-neighbor interaction and the presence of a next-nearest-neighbor interaction are both sources of frustration and open the possibility to have a quite rich phase diagram for this model. This model is suitable to describe cuprates with edge-sharing $CuO_2$ plaquettes where the bonding angle between two nearest Coppers and the intermediate Oxygen is slightly larger than $90^\circ$, resulting in a ferromagnetic nearest-neighbor interaction term with an intensity comparable to the antiferromagnetic next-nearest-neighbor interaction term. According to this, the analysis of the phase diagram of such a model is relevant not only on the pure theoretical level (effects of frustration, incommensurability, spiral ordering, dimerization), but also on the level of understanding real materials and their applications.

The manuscript is organized as follows. In the next section, we present the Hamiltonian under study and give few details about the numerical framework within which the model has been solved. In section three, we describe the entanglement measure we have chosen to compute and give the reasons behind such a choice. The results of the analysis are discussed in section four. Finally, we draw some conclusions and give some perspectives.

\section{Model and Method} \label{MM}

The Hamiltonian under analysis reads as:
\begin{multline}
H = -J_z \sum_i S^z_i S^z_{i+1}  \\ + J_{\bot}
\sum_i ( S^x_i S^x_{i+1} + S^y_i S^y_{i+1})
+  J^{\prime}\sum_i \mathbf{S}_i \mathbf{S}_{i+2}
\label{ham}
\end{multline}
where $J_z>0$ is the ferromagnetic z-axis nearest-neighbor coupling constant, $J_{\bot}>0$ is the antiferromagnetic in-plane nearest-neighbor coupling constant and $J^{\prime}>0$ is the antiferromagnetic isotropic next-nearest-neighbor coupling constant.

We solved the Hamiltonian (\ref{ham}) numerically by means of the DMRG~\cite{dmrg} technique on a chain with 100 sites. DMRG forced us to use open-boundary conditions. We have retained up to 200 states per block at every step of the renormalization procedure. The finite-size effects, enhanced by the open-boundary conditions, have been systematically mitigated by taking into account, in the average procedures, only the central part of the system, i.e. by neglecting the contributions coming from the sites close to the edges.

It is worth noticing that the region of model-parameter space we have explored is just inaccessible to any of the almost exact field theories as no small parameter can be easily identified. Only a very powerful numerical technique, such as DMRG, would be capable to bridge the gap between the few known exact results for this model and to provide reliable reference data.

\section{Concurrence}

There exist a few entanglement measures, which mainly differ in the way the system is split into two blocks whose entanglement is measured: a reference block, whose properties (averages and correlation functions) will come into play, and the rest of the system, which will simply act as a bath. In spin systems, the one-tangle (i.e., when we choose as reference system a single spin), or von Neumann entropy, is a function of the local magnetization only and, hence, is no more informative than this latter. Then, in such systems, in order to catch some more physics than only the one related to the ferromagnetic phase, it is necessary to use the concurrence~\cite{wootters}, or pairwise entanglement (i.e., the reference system will now be two spins), which depends on spin-spin correlation functions.

The concurrence for a couple of spins residing at sites $i$ and $j$, respectively, is defined as:
\begin{equation}
C_{i,j} = \textrm{max}( 0, \lambda_1 - \lambda_2 - \lambda_3 - \lambda_4 )
\label{def_conc}
\end{equation}
where $\{\lambda_i\}$ are the eigenvalues, in decreasing order, of a positively definite matrix $R$ defined as follows:
\begin{equation}
R=\sqrt{\rho(\sigma^y \otimes \sigma^y)\rho^{*}(\sigma^y \otimes \sigma^y)}
\end{equation}
where $\sigma^y$ is just the second Pauli matrix and $\rho$ is the reduced density matrix. This latter can be computed by integrating out, in the ordinary density matrix of the system, all degrees of freedom except for those of the two spins under analysis.

If one integrates out the degrees of freedom of the bath analytically, the reduced density matrix of two spins reads as:
\begin{equation}
\rho_{ij}=\frac{1}{4}
\left(
\small
\begin{array}{cccc}
1 + K^{zz}_{ij} +2 m^z & 0& 0& 0\\
0& 1 - K^{zz}_{ij} &2 K^{xx} &0 \\
0& 2 K^{xx} &1 - K^{zz}_{ij} &0 \\
0& 0& 0& 1 + K^{zz}_{ij} - 2 m^z
\end{array}
\right)
\end{equation}
where $K^{zz}_{ij}=4 \langle S^z_i S^z_j \rangle$, $K^{xx}_{ij}=4 \langle S^x_i S^x_j \rangle$, and $m^z=2 \langle S^z_i \rangle$. We have assumed that there is no anisotropy in the $x-y$ plane. The extreme values of the concurrence, zero and one, indicate that the system is either a product state or a maximally entangled one, respectively. Wootters~\cite{wootters} demonstrated that the concurrence can be directly related to the entropy of formation for two spins $1/2$ both for pure and mixed states.

In order to take into account the contribution to the entanglement coming from the correlations at all distances, in this manuscript, we have adopted $\tau_2$ \cite{coffman} as the reference entanglement measure:
\begin{equation}
\tau_2 = \sqrt{\sum_{d > 0} C^2_{i,i+d}}
\label{tau2}
\end{equation}
According to our average procedure (see Sec.~\ref{MM}), $C_{i,i+d}$ does not depend on $i$ and so does $\tau_2$.

\section{Results}

\begin{figure}
\includegraphics[width=0.48\textwidth]{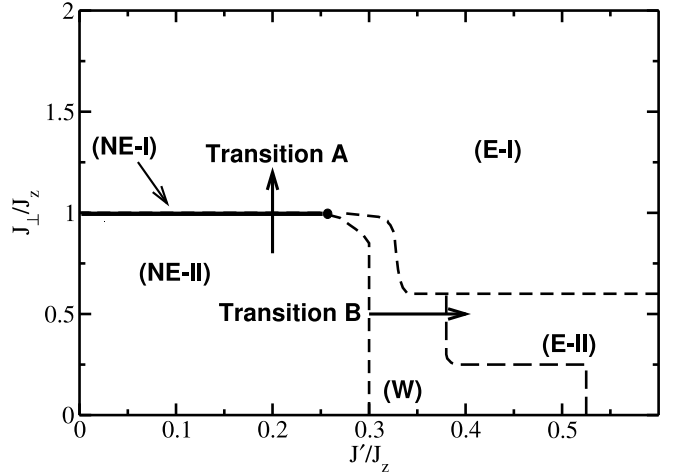}
\caption{Schematic phase diagram of the model (\ref{ham}). See Ref.~\protect\cite{ours} for a detailed description of the phases.} \label{fig1}
\end{figure}

\begin{figure*}
\includegraphics[angle=-90,width=0.48\textwidth]{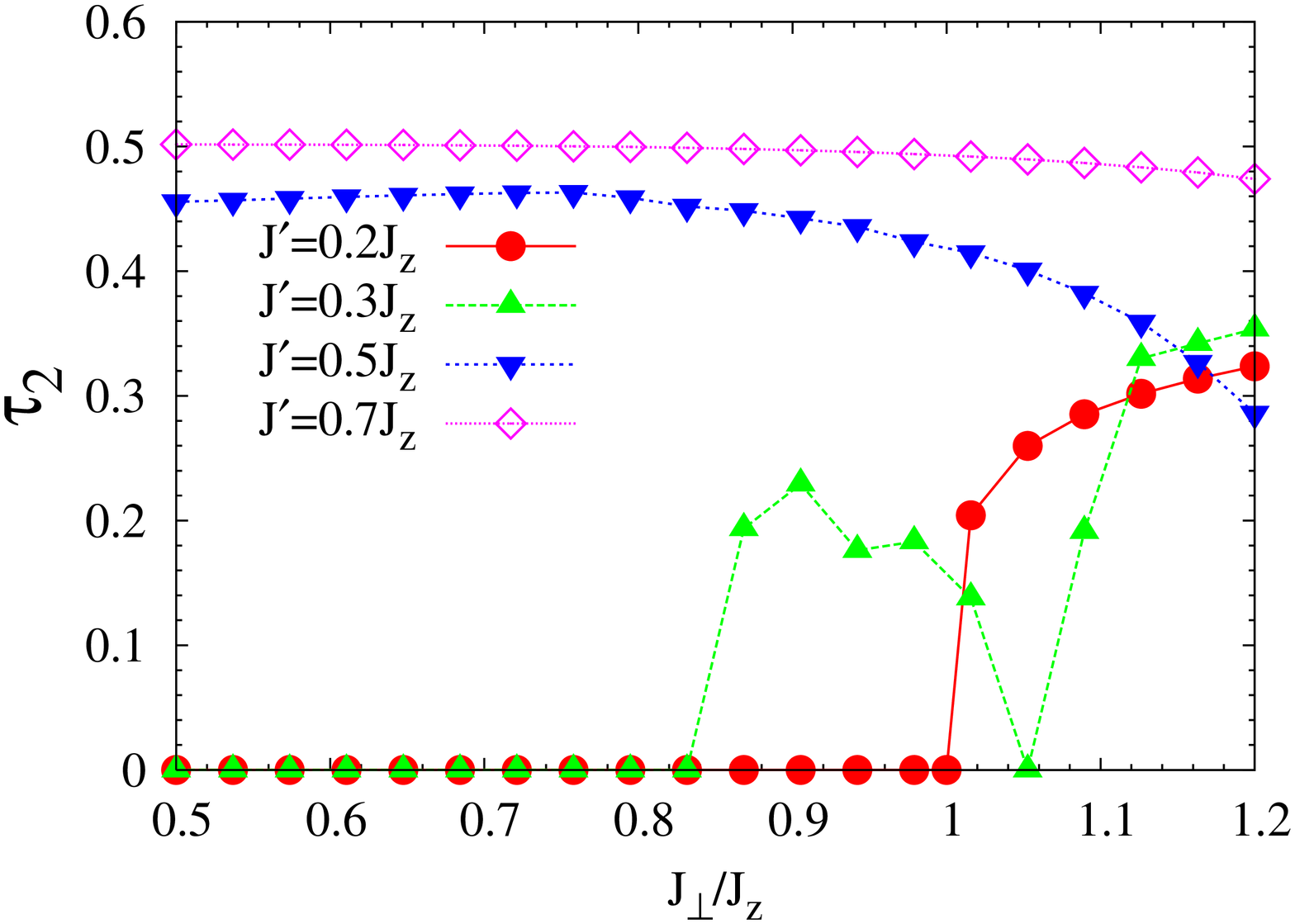}
\includegraphics[angle=-90,width=0.48\textwidth]{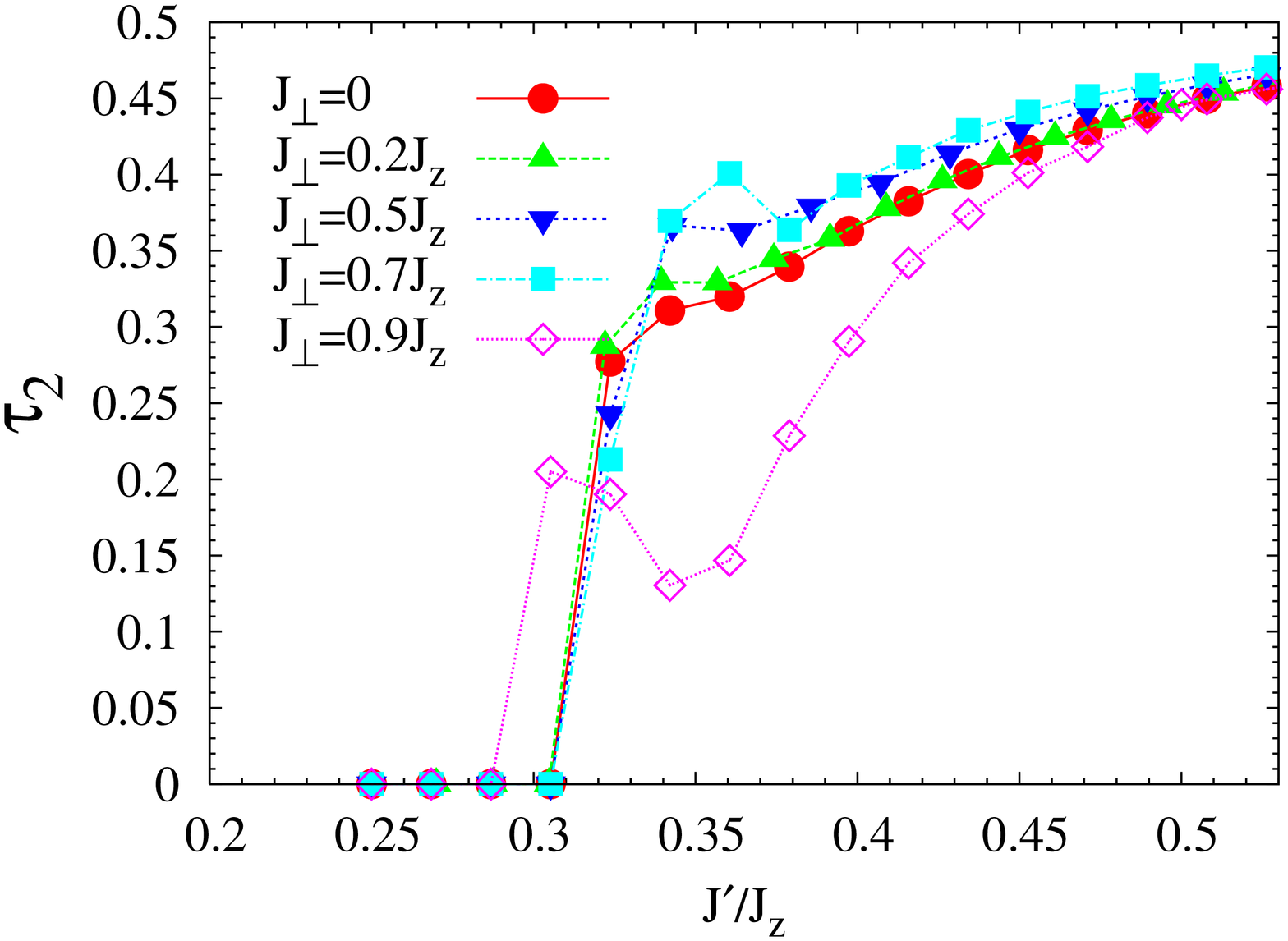}
\caption{$\tau_2$ as a function of (left) $J_\bot/J_z$ and (right) $J'/J_z$.}
\label{fig2}
\end{figure*}

The starting point of our analysis is the doubly degenerate, completely polarized state, located within the phase diagram of the system in the region $J_{\perp} < J_z$ --- $J^{\prime} \lesssim 0.31 J_z$, as found in Ref.~\cite{ours} (see Fig.~\ref{fig1}).

First, we analyze the behavior of $\tau_2$ across the transition of the A type (see Fig.~\ref{fig1}). At $J_{\perp} \gg J_z$ and $J^{\prime}=0$, the model reduces to the exactly solvable $XX$ model which shows an in-plane antiferromagnetic quasi-long-range order. We do expect an $XX$-model-like behavior all the way down to $J_{\perp} \gtrsim J_z$ \cite{ours_kosice}. As can be seen from Fig.~\ref{fig2} (left), $\tau_2$ successfully detects the increase of complexity of the ground state trough the A-type phase transition. At $J^{\prime}=0.3J_z$, a region with non-zero entanglement is present well below the isotropic line ($0.82J_z<J_{\perp}<1.05J_z$) and can be interpreted as the appearance of a third phase between the ferromagnetic and the $XX$-model-like ones. For larger values of $J^\prime$, in the range of $J_\perp$ values explored by us, $\tau_2$ is finite, but almost featureless. Entanglement measures other than $\tau_2$, such as measures involving more than two spins in the reference system, could reveal the presence of other phase transitions in this region. In particular, entanglement measures involving four spins should be able to check the tendency towards dimerization. In A-type transitions, $C_{d}$ is non zero only for values of $d$ up to $15$ lattice spacings.

Then, we examine the features of $\tau_2$ as $J^{\prime}$ exceeds the critical value of approximately $0.31 J_z$ for $J_{\perp}<J_z$ (transitions of B type on Fig.~\ref{fig1}). The increasing frustration induced by the next-nearest-neighbor term steadily reduces the ferromagnetic polarization of the spins as the intensity of the antiferromagnetic correlations between next-nearest-neighbors increases. As a matter of fact, within the region $0.31J_z < J^{\prime} < 0.4J_z$, we have found a finite magnetization per site together with an increasing next-nearest-neighbor antiferromagnetic correlation length. From an entanglement point of view, a completely polarized ferromagnetic state has zero concurrence since it is a product state. Therefore, $\tau_2$ is expected to increase from zero to a finite value across the transition. Indeed, such a behavior was already observed in our previous Lanczos calculations~\cite{ours_houston} on a 24-site system, but the quite relevant finite-size effects led to the appearance of steps in $\tau_2(J^{\prime})$ that mined our comprehension of the order and nature of the transition. The current DMRG calculations on a 100-site system are not affected by such drawbacks. It can be very clearly seen in Fig.~\ref{fig2} (right) that $\tau_2$ is quite sensible with respect to this transition. Up to values of $J_{\perp}=0.7J_z$, $\tau_2$ is almost independent on $J_\perp$ and presents only a wide peak immediately after the transition. However, for $J_{\perp}=0.9J_z$, this peak evolves into a pronounced maximum at $0.3J_z<J^{\prime}<0.35J_z$, which, together with the above noted analogous increase at $0.82J_z<J_{\perp}<1.05J_z$ and $J^{\prime}=0.3J_z$, indicates the presence of a well defined region in the parameter space that is a good candidate to be recognized as a new ordered phase of the system. The nature of the ordering ruling such a phase can be deeply understood only by studying the spin-spin correlation functions and such a work is currently in progress. It is worth noting that our present calculations confirm an earlier observation that the only non-zero contributions to $\tau_2$, for $J_\perp<0.9 J_z$ in B-type of transitions, are those coming from second-neighbor spin-spin correlation functions. For $J_\perp=0.9 J_z$, we have found that also third-neighbor spin-spin correlations contribute to $\tau_2$.

\section{Conclusions}

We have studied a 100-site anisotropic extended F-AF Heisenberg chain by means of Density Matrix Renormalization Group retaining 200 states per block at every renormalization stage. We have measured the concurrence at all distances and checked its capability to detect phase transitions leaving the fully polarized ferromagnetic phase of the system on varying the frustration driven by the anisotropy in the nearest-neighbor coupling and by the presence of next-nearest neighbor coupling. Although it was not possible to establish neither the order of the transitions nor the nature of the newly appearing phases, the behavior of the concurrence clearly showed their presence. Moreover, by means of the analysis of concurrence features, we have been able to identify a well defined region in the parameter space that shows the signatures of a complex spin ordering. In order to clarify the nature of this probable new phase, we have just started, and it is still in progress, the analysis of spin-spin correlation functions in this region.

\end{document}